\documentclass[printer]{aa}
\usepackage{graphicx}
\usepackage{txfonts}
\usepackage{color}
\usepackage[dvipsnames]{xcolor}
\usepackage{float}
\usepackage{lscape}
\usepackage[colorlinks=true,linkcolor=blue,filecolor=blue,citecolor=blue,urlcolor=blue, draft=False]{hyperref}
\usepackage{natbib}
\usepackage{multirow}
\usepackage{amstext}
\usepackage{siunitx}
\usepackage{caption}
\DeclareCaptionFormat{cont}{#1#2#3\par}

\newcommand{\citel}[1]{\citeauthor{#1}\,\citeyear{#1}}
\newcommand{\Msun}[1]{\,M$_{\odot}$}
\newcommand{\kmpersec}[1]{\,km.s$^{-1}$}

\begin{document}

\titlerunning{The detectability of high-z GRB near-infrared afterglows with CAGIRE}
   \authorrunning{F.\,Fortin et al.}

   \title{The detectability of high-redshift gamma-ray bursts near-infrared afterglows with CAGIRE}

   \author{
           F. Fortin\inst1 \and
           J.L. Atteia\inst1 \and
           A. Nouvel de la Fl\`eche\inst1 \and
           H. Valentin \inst1 \and
           O. Boulade\inst2 \and
           D. Corre\inst2 \and
           D. Turpin\inst2 \and
           A. Secroun\inst3 \and
           S. Basa\inst4 \and
           F. Dolon\inst4 \and
           J. Floriot\inst4 \and
           S. Lombardo\inst4 \and
           J. F. Le Borgne\inst1$^,$\inst4 \and
           A. M. Watson\inst5 \and 
           W. H. Lee\inst5           
          }

   \institute{IRAP, Universit\'e de Toulouse, CNRS, CNES, UPS, 31401 Toulouse, France\and
   CEA-IRFU, Orme des Merisiers, 91190 Gif-sur-Yvette, France.\and
   Aix Marseille Universit\'e, CNRS/IN2P3, CPPM, Marseille, France\and
   LAM, Universit\'e Aix-Marseille \& CNRS, UMR7326, 38 rue F. Joliot-Curie, 13388
Marseille Cedex 13, France\and
   Instituto de Astronom\'ia, Universidad Nacional Autonoma de M\'exico, Apartado Postal 70-264, 04510 M\'exico , CDMX, Mexico}

   \date{received ... ; accepted ...}

 
  \abstract
    {
    Transient sky astronomy is entering a new era with the advent of the \textit{SVOM} mission (Space Variable Objects Monitor), which was successfully launched on the 26$^{th}$ of June, 2024. The primary goal of \textit{SVOM} is to monitor the hard X-ray sky searching for gamma-ray bursts (GRBs). On top of its on-board follow-up capabilities, \textit{SVOM} will be backed by its ground segment composed of several facilities, of which the near-infrared imager CAGIRE. Mounted on the robotic telescope COLIBRI, it will be a unique instrument, able to perform fast follow-up of GRB afterglows in J and H bands, an ideal combination to catch high-redshift (z$>$6) and/or obscured GRBs.
    }
    {
    This paper aims at estimating the performances of CAGIRE for GRB near-infrared afterglow detection based on the characteristics of the detector and the specificities of the COLIBRI telescope. Quickly fading GRB afterglows pose challenges that should be addressed by adapting observing strategies to the capabilities of CAGIRE.
    }
    {
    We use an end-to-end image simulator to produce realistic CAGIRE images, taking into account results from the characterization of the ALFA detector used by CAGIRE. We implemented a GRB afterglow generator that simulates infrared lightcurves and spectra based on published observation of distant GRBs (z$>$6).
    }
    {
    We retrieved the photometry of 9 GRB afterglows in various scenarios covered by CAGIRE. Catching afterglows as early as two minutes after burst allows the identification of a nIR counterpart in the brightest 4 events. When artificially redshifted even further away, these events remain detectable by CAGIRE up to z=9.6 in J band, and z=13.3 in H band, indicating the potential of CAGIRE to be a pioneer in the identification of the most distant GRBs to date.
    }
    {}


   \maketitle
%

\section{Introduction}\label{sect:introduction}
Gamma-ray bursts (GRBs) are high-energy transient events that intervene during the evolution of massive stars. They are generally classified according to the duration of the burst (\citel{1992AIPC..265..304D}, \citel{1993ApJ...413L.101K}); short GRBs (t$_{90}$<2\,s) are confirmed since 2017 to originate from the merger of neutron star binaries (BNS, \citel{2017ApJ...848L..13A}), while long GRBs (t$_{90}$>2\,s) have been associated to the collapse of massive stars \citep{1999ApJ...524..262M}. Both supernovae (SN, \citel{2006ARA&A..44..507W}) and kilonovae (KN, \citel{2017ApJ...848L..12A}) events have been associated to GRBs. All GRBs are described by a prompt emission of high-energy photons, followed by an afterglow that quickly fades on the scale of several hours. The prompt emission is produced close to the central engine of the GRB, likely inside the jets outflowing from a newly formed black hole (or magnetar, see e.g. \citel{2011MNRAS.413.2031M}, \citeyear{2017ApJ...841...14M}). The afterglow comes from the interaction of the jets with the surrounding medium, and can radiate from hard X-rays down to radio wavelengths.

The observation of GRBs provides an ideal window into black hole physics, the accretion-ejection mechanisms that lead to the formation of relativistic jets and the acceleration of cosmic rays, and more broadly into the evolution of massive stars with the association of GRBs to SN as well as BNS gravitational mergers. The intrinsic luminosity of GRBs allows some of the brightest events to be seen at high redshifts (z > 6), hence they can be used to probe the conditions of the early universe and constrain the evolution of massive star formation, the composition of the host galaxies or even of the intergalactic medium. The first generation of stars, the population III stars, are also reachable at such redshifts. However, the followup of far away GRB afterglows becomes impossible even with fast-reacting optical observatories past z$\sim$6, as radiation below the Lyman\,$\alpha$ wavelength is progressively shifted out of the optical domain as redshift increases.

The \textit{SVOM} mission (Space Variable Objects Monitor, \citel{2016arXiv161006892W}) is a space-based GRB observatory with both on-board and ground follow-up capabilities. The main high-energy instrument that will detect the prompt emission of GRBs, ECLAIRs, will trigger automatic follow-up on the ground segment, which is composed of a collection of small robotic telescopes (GWAC) as well as two 1\,m class telescopes, the C-GFT and the F-GFT. The latter, also known as COLIBRI \citep{2022SPIE12182E..1SB}, is a 1.3\,m robotic telescope installed in San Pedro Mart\'ir observatory, Mexico. It will be capable of observing 3 different channels simultaneously: two channels with the optical imager DDRAGO \citep{2022SPIE12184E..7WL} operating in the (g, r, i) and (z, y) bands, and one channel with the near-infrared (nIR) imager CAGIRE (CApturing Gamma-ray burst InfraRed Emission, \citel{cagiresimulator}) operating in the J and H bands. DDRAGO is designed to have its 4k$\times$4k detectors cover a field of view (FoV) of 26\arcmin, which will fully cover CAGIRE's 21.7\arcmin\, FoV. DDRAGO will be operational by the end of 2024; once CAGIRE is operational (starting summer 2025), the combination of both instruments will make possible the early sampling of afterglow spectral energy distributions over the optical to near-infrared domain, which in turn will allow to determine photometric redshifts in the range z = 3.5\,--\,8 with 10\% accuracy \citep{2018SPIE10705E..1RC}.

Mounted on the Nasmyth focus of COLIBRI, CAGIRE will be able to meet the observational challenges raised by far away and obscured GRBs. The camera is capable of J and H band photometry to probe sources beyond redshift 6, and its field of view was designed to cover most --if not all-- of the error boxes associated to the localization of GRBs with ECLAIRs. COLIBRI will allow CAGIRE to begin imaging as soon as one minute after the initial trigger from ECLAIRs (see SVOM System Requirements Document 0702), which is ideal to catch the early times of the fast-decaying afterglow lightcurves. This combination of COLIBRI's quick reactivity with CAGIRE's nIR imaging capabilities is unique in the field of transient sky astronomy.

In this paper, we aim to provide insights into the future capabilities of CAGIRE by analysing its expected performance on already observed distant GRBs detected in the last two decades. We first provide the characteristics of CAGIRE and describe the softwares we used for image simulation and data processing (Section\,\ref{sect:cagire}). We then build a sample of modelled lightcurves from past GRB afterglows in Section\,\ref{sect:grbDB} that we then feed into the CAGIRE image simulator to test their detectability in various realistic scenarios (Section\,\ref{sect:results}). We then discuss those results and conclude in Sections\,\ref{sect:discussion} and \ref{sect:conclusion}.

\section{CAGIRE: instrument, image simulation and signal extraction}\label{sect:cagire}

\subsection{Instrument description and specifications}

CAGIRE is a nIR imager that operates in the J (1.22\,$\mu$m) and H (1.63\,$\mu$m) bands. The detector is a 2k$\times$2k Astronomical Large Format Array (ALFA) loaned by ESA, sampling the sky at 0.65\arcsec/pix which covers an FoV of 21.7\arcmin. The camera is cooled down to operate at 100\,K, which allows for a negligible dark current (0.004\,e$^-$/s/pix) and for CAGIRE to be sky-limited. The ALFA detector can be read in a non-destructive manner, so that CAGIRE will generate ramps composed of several frames each with a fixed exposure of 1.33\,s. The edges of the detector are constituted of a ring of pixels that are not sensitive to light, and serve as reference pixels to correct for common mode electronic noise.

\subsection{CAGIRE image simulator}
We performed simulations of images acquired by CAGIRE at the Nasmyth focus of COLIBRI using the dedicated software described in \cite{cagiresimulator}. The simulator was initially based on the software developed by D. Corre \citep{2018SPIE10705E..1RC} dedicated to simulating the output images of the COLIBRI telescope. \cite{cagiresimulator} adapted the image simulation and the exposure time calculator to work with CAGIRE and its specificities. The sources within the simulated field of view are taken from the 2MASS catalogue \citep{2003yCat.2246....0C} and the sky background added depending on the photometric band. The simulator also includes random cosmic rays hitting the detector during exposures. CAGIRE-specific effects are then applied according to preliminary characterization of the detector done at CEA and CPPM: the flat field, dark current, hot and cold pixels, as well as the persistence of pixels from previous acquisition (in the case they were filled up to at least 80\% of their saturation level). Interpixel cross-talk and the non-linearity from the capacitance effect are added before finally applying the offset level of the detector.

The simulator is able to generate a fading GRB afterglow based on tabulated data included with the software, with lightcurves sampled at the CAGIRE frequency of $\sim$1.33\,s/frame. The base version of this software is publicly available on GitHub\footnote{\url{https://github.com/alixdelafleche/simu}}, however it is no longer maintained.

For this study, we have worked on improving the performances of the CAGIRE image simulator and implementing additional features. Several routines have been parallelised and optimised to be able to generate ramps much quicker, up to an average rate of 2.5 frames per second for longer exposures. The output ramps of the simulator follow the same \textsc{.fits} data and header formatting as the real instrument. The main addition compared to the base version is the presence of a GRB generator, which uses models of afterglow lightcurves that are input in the simulator to recreate a fading transient in the final images. It is possible to simulate an afterglow at any time after burst, in both J and H bands, as well as to change the redshift of the simulated GRB (see more details in Section\,\ref{sect:grbDB}). This new version of the CAGIRE image simulator is developed in a private GitLab repository hosted by the Laboratoire d'Astrophysique de Marseille, and is mirrored in a public GitHub repository\footnote{\url{https://github.com/COLIBRI-CAGIRE/CagireImageSimulator}}.

\subsection{Observing conditions}\label{subsect:obs_cond}
As GRBs are unpredictable transients and require quick follow-up, COLIBRI will not wait for optimal observing conditions after an alert issued by ECLAIRs. As such, we consider in the rest of the paper that observations are made in sub-optimal conditions. The seeing at zenith is set to 1\arcsec, and the elevation of the targets at 41.8\degr\, corresponding to an airmass of 1.5. The moon age is fixed at 7 days after new moon, although the moon phase has little impact on the sky brightness in J and H bands. The sky brightness in San Pedro Mart\'ir is measured to be 17\,mag/arcsec$^2$ in J band and 15\,mag/arcsec$^2$ in H band \citep{watson2018} normalized for an airmass of 1.5.

\subsection{Pre-processing of CAGIRE images}
As CAGIRE will operate in up-the-ramp mode, the raw output files consist of a series of frames with increasing signal. The goal of the pre-processing software is to convert the raw ramps into a single image containing the flux in ADU/frame, corrected for instrumental effects, which are ready to be fed into any astrophysical pipeline for data processing (i.e. astrometry, photometry, search for transients). The main tasks of the pre-processing software include the correction of common mode electronic noise and offset of the detector using the ring of reference pixels around the ALFA detector, the creation of differential ramps and their fitting to retrieve the flux for each pixel. The software automatically detects when and where the detector was impacted by a cosmic ray, and whether or not pixels attained their saturation level during the exposure; the fitting range of the flux is then adjusted accordingly.

A Python version of the pre-processing software is publicly available on the COLIBRI-CAGIRE GitHub collaboration\footnote{\url{https://github.com/COLIBRI-CAGIRE/CagirePreproc}}. It can be used to process both real CAGIRE data and images output from the CAGIRE Image Simulator.

\subsection{Extraction of the photometry}
We determine the zero point of the photometry for each simulated image using the stars available in the field of view (N$\sim$1000), which are generated along with the GRB afterglow using catalogued magnitudes that we retrieve through the CDS service Simbad \citep{2000A&AS..143....9W}. We extract their flux by performing PSF fitting photometry with the \textsc{Python/photutils} package. The fitting of a symmetrical two-dimensional gaussian PSF is performed in a 15 pixels-wide window around the position of the sources. The background is estimated locally and assumed to be a constant in the fitting window.

The zero points are obtained by fitting the instrumental magnitudes against the catalogued magnitudes. We measure average zero points of ZP$_J$=20.90$\pm$0.03 mag and ZP$_H$=21.26$\pm$0.04 mag in the observing conditions described in Section\,\ref{subsect:obs_cond}. We ensured that the values of the zero points were not affected by the brightness of the stars, as the PSF may deviate from a gaussian for either very bright or very dim sources.

\subsection{Limiting magnitude of CAGIRE}
We derive the limiting magnitude at a given signal-to-noise ratio (SNR) of each image using the generic formula linking the magnitude of a given reference star to its measured SNR:
\begin{equation}
    \mathrm{Mag}_{lim, SNR} = \mathrm{Mag}_{ref} -2.5\,\mathrm{log}_{10}\left( \frac{\mathrm{SNR}}{\mathrm{SNR}_{ref}}\right)
\end{equation}

The measure of the SNR is performed on each reference star with a magnitude comprised between 14 and 16.5, which is the range in brightness where stars do not approach saturation and are still sufficiently bright to extract precise photometry using PSF fitting. The noise from the sky and detector readout is estimated locally in an annulus around each star. We estimate the general detection limit using the median of those measurements.

\begin{figure}
    \centering
    \includegraphics[width=.85\columnwidth]{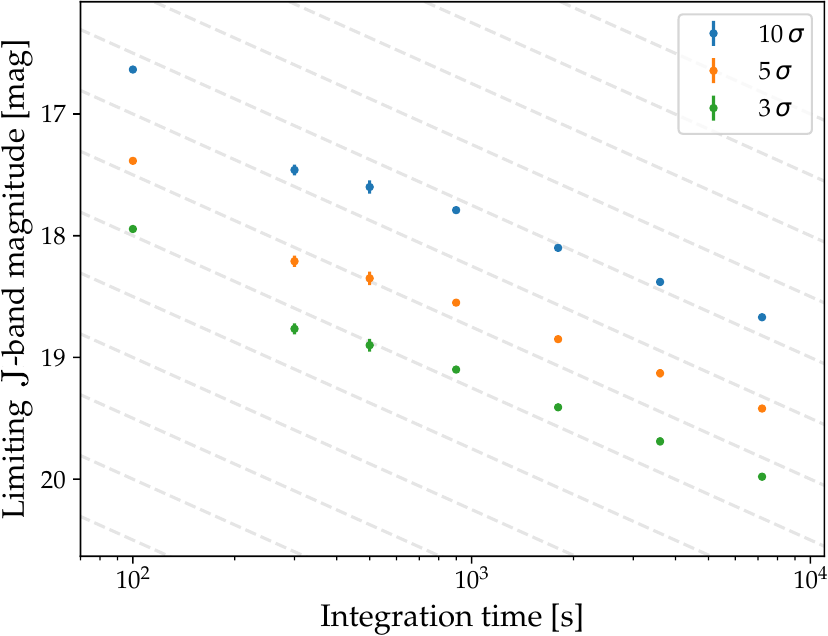}
    \includegraphics[width=.85\columnwidth]{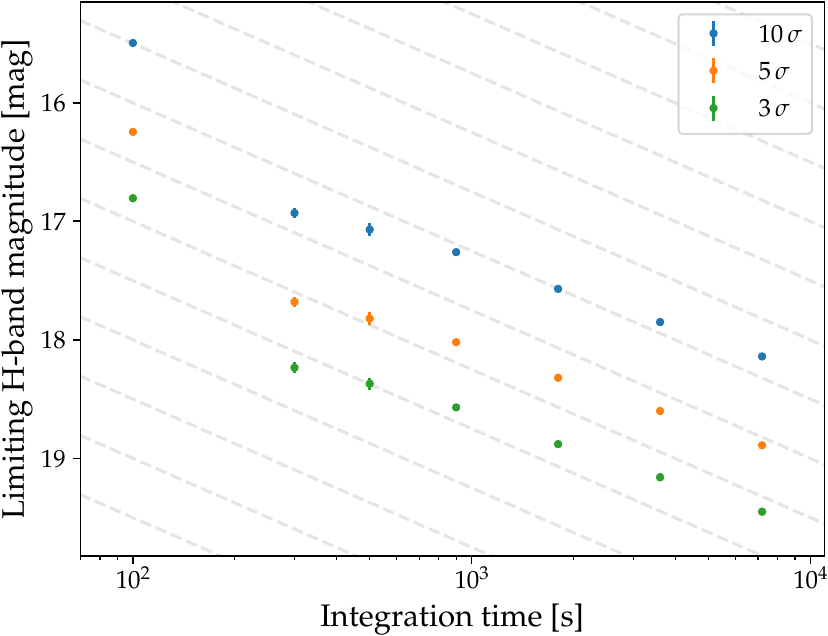}
    \caption{Detection limits of CAGIRE. The exposure times are split into 100\,s (75 frames) individual exposures to avoid detector saturation due to the sky background. Grey dashed lines indicate the maximum theoretical evolution of the limiting magnitude (SNR increasing with the square-root of exposure).}
    \label{fig:detlim}
\end{figure}

We present the limiting magnitudes computed for a SNR of 10, 5 and 3 for various exposures in Figure\,\ref{fig:detlim}. The typical error on the limiting magnitudes is around 0.05 mag. We work with individual exposures of 100\,s in both bands, since the sky contribution in H band will already fill more than half of the dynamic range of the detector for this integration time. Exposures longer than 100\,s are split into 100\,s individual exposures taken in a dithering pattern, i.e. randomly offset on the sky by 30\arcsec, which are realigned and averaged afterwards. The sky contribution is subtracted using the median of non-aligned images. We note that the drastic evolution of the limiting magnitude between a single 100\,s exposure and three 100\,s exposures in a dithering pattern is mostly due to the bad subtraction of the background sky and detector defects on the single exposure. The evolution of the limiting magnitude for total exposure times equal or greater than 300\,s follows the expected trend (SNR increasing with the square-root of total exposure time), albeit  with a shallower slope due to the extra readout noise, which is partly mitigated as dithering frames increase in number, up to 72 frames of 100\,s for a two hour exposure.


We also explored the possibility of increasing the individual exposure time in J band only, as the sky contribution is much lower than in H band. We increased the individual J band exposures up to 300\,s (225 frames), point at which the pre-processing time becomes longer than the acquisition time. We obtain a detection limit for 3$\times$300\,s (900\,s total) in the J band at 19.7 (SNR=3), compared to 19.1 for 9$\times$100\,s. The increased detection limit is due to the fewer readouts of the detector, as well as the use of a higher number of frames for the fit on the differential ramps, which provides more constrained value for the flux. Note that this advantage towards greater exposures times in J band is mitigated in the case of quickly fading sources such as GRB afterglows, which we will discuss in Section\,\ref{sect:results}.

We will now use the simulated images from CAGIRE to evaluate the detectability of GRB afterglows at high redshift (z$>$6) in J and H bands. We first present how we model the afterglow lightcurves, and then explore various observational scenarios depending on the delay between the bursts and the first images, as well as the redshift of the events.

\section{Gamma-ray burst database, generating lightcurves and spectra}\label{sect:grbDB}

We compiled modelled infrared lightcurves of GRB afterglows from the literature for events located at redshift 6 and above. We collected a total of 9 GRBs with sufficient coverage of their afterglow (see Table\,\ref{tab:grb_prop}) to model the evolution of their lightcurve in the observer's rest frame. The temporal evolution of the lightcurves is modelled using a decaying broken powerlaw that, depending on each GRB, may present a break starting from several minutes up to a couple of days after the initial burst alert, and is described by:
\begin{equation}
    G(t) = \left( \frac{t}{t_{b}} \right)^{-\alpha_1} \left[ \frac{1}{2} \left(1+\left( \frac{t}{t_b}\right)^{\frac{1}{\delta}}\right) \right]^{\delta (\alpha_1 - \alpha2)} 
\end{equation}
with $\alpha_1$ and $\alpha_2$ the decay indices before and after the break occurring at t$_b$, and $\delta$ a parameter that governs the smoothness of the transition between the two regimes. For the sake of simplicity, we consider the transition to be instantaneous, hence we fix $\delta << $1. 

\begin{figure}[h]
    \centering
    \includegraphics[width=0.99\columnwidth]{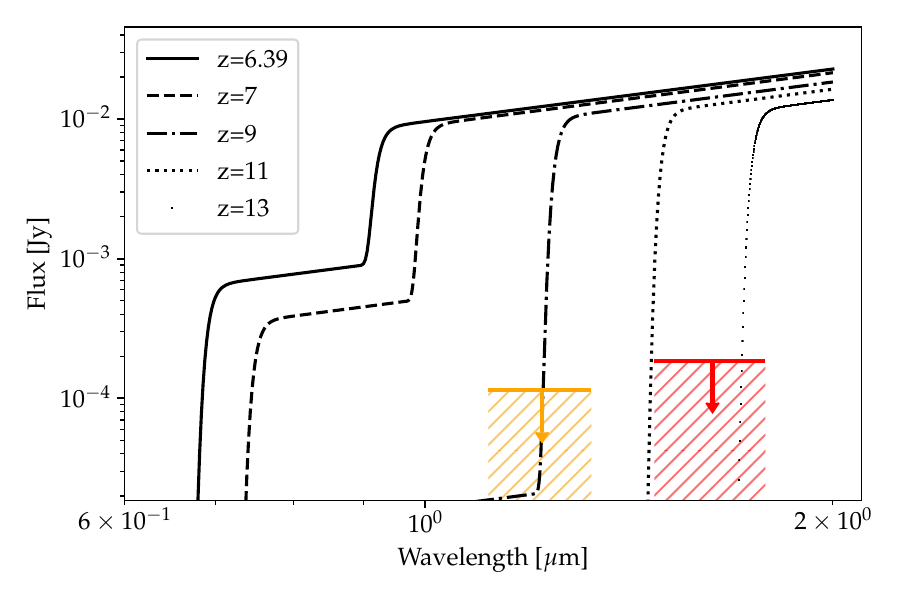}
    \caption{Optical to near-infrared model spectrum of GRB 050904A at the midpoint of 300\,s observations started 1\,min after burst. The corresponding CAGIRE detection limits at SNR=3 are displayed for the J-band (orange) and H-band (red).}
    \label{fig:grb050904spec}
\end{figure}

The spectral shape of the afterglows is considered constant over time, and is modelled by a powerlaw:

\begin{equation}
    H(\nu) = \nu^{-\beta}
\end{equation}
with $\beta$ the spectral index. A cut-off is introduced above $\nu_{Ly\alpha}$=2.47$\times$10$^{15}$\,Hz, the starting point of the Lyman forest absorption. Between the Lyman\,$\alpha$ and Lyman\,$\beta$ transitions, the loss of flux is calculated according to the relation given in \cite{1993ApJ...418..601Z}, which scales the amount of absorbers with redshift. We note that this is an extrapolation beyond z$\sim$4; for reference, according to this model, the loss is of 82\% at redshift 6 (between 700--850\,nm in the observer's frame), and 98.5\% at redshift 8 (between 900--1090\,nm). Figure\,\ref{fig:grb050904spec} provides modelled spectra of GRB 050904A to illustrate this evolution at various redshifts.

The flux of the GRB afterglow is thus governed by:

\begin{equation}
    F(\nu,t) \propto G(t) \times   H(\nu)
\end{equation}

The lightcurves are normalized using the first measured magnitude in each observation band (listed in Table\,\ref{tab:grb_prop}). It is then possible to recover the afterglow lightcurves in the rest frame of the GRB. To do this, we take into account the effects of luminosity distance, spectral redshift and time dilation, all three following the equations 6 and 7 given in \cite{2000ApJ...536....1L}. We finally apply the same equations backwards to artificially place observed GRBs as if they were located further away, at higher redshifts.

\begin{table*}[h]
    \small
    \caption{General information on high-redshift GRBs and model parameters of their infrared afterglow.}
    \label{tab:grb_prop}
    \begin{tabular}{lllllllllllll}
    \hline\hline\\[-2ex]
        Identifier  & z & R.A. & Dec. & T$_0$ & T$_J$ & T$_H$ & m$_J$ & m$_H$ & $\alpha^{(a)}$ & t$_{break}$ & $\beta^{(b)}$ & Ref. \\
         & & [deg] & [deg] & [MJD-53000] & [hr] & [hr] & [mag] & [mag] & & [d] & & \\
    \hline\\[-2ex]
        GRB 050904A & 6.39  &  13.7116 &  14.0860 &  617.077593 & 3.07 & 9.79 & 17.36 & 18.17 & (0.9,1.9)     & 2.1  & 1.25 & [1] \\
        GRB 080913A & 6.695 &  65.7280 & -24.1295 & 1722.282570 & 0.10 & 0.10 & 19.96 & 19.64 & 1.03          & -\,- & 1.12 & [2] \\
        GRB 090429B & 9.4   & 210.6667 &  32.1706 & 1950.2292   & 2.95 & 3.27 & 22.8  & 21.41 & (0, 0.53)     & 0.02 & 0.51 & [3] \\
        GRB 090709A & 8.5   & 289.9277 &  60.7276 & 2021.318449 & 0.11 & 0.03 & 18.43 & 15.52 & 0.9           & -\,- & 3.8  & [4] \\
        GRB 100205A & <8    & 141.3875 &  31.7403 & 2232.179664 & 5.48 & 4.73 & 24.29 & 23.63 & 0.43         & -\,- & 0.51 & [5] \\
        GRB 120521C & 6.03  & 214.2833 &  42.1431 & 3068.973692 & 7.63 & 8.54 & 20.4  & 19.8  & (-0.83, 1.38) & 0.34 & 0.34 & [6] \\
        GRB 120923A & 7.8   & 303.7958 &   6.2203 & 3193.219514 & 1.37 & 3.24 & 21.9  & 21.4  & (0.25, 2)     & 1.46 & 0.39 & [7] \\
        GRB 140515A & 6.32  & 186.0639 &  15.1046 & 3792.38375  & 15.6 & 14.54 & 20.63 & 20.61 & 0.89          & -\,- & 0.33 & [8] \\
        GRB 210905A & 6.318 & 309.0481 & -44.4396 & 6462.0088   & 0.36 & 0.13 & 17.12 & 16.15 & (0.69, 0.94)  & 0.99 & 0.6  & [9] \\  
    \hline\\[-2ex]
    \end{tabular}
    \tablefoot{The first magnitude measurements in J and H bands are given along with the time after burst they were performed.}
    $^a$: temporal decay index of the GRB as measured in the frame of the observer. If two values are provided, an associated time break is given ($t_{break}$).
    
    $^b$: spectral index of the GRB as measured in the frame of the observer. Is assumed to be constant over time.
    \tablebib{
    [1]\,\cite{2006Natur.440..181H}; [2]\,\cite{2009ApJ...693.1610G}; [3]\,\cite{2011ApJ...736....7C}; [4]\,\cite{2010AJ....140..224C}; [5];\,\cite{2019MNRAS.488..902C}; [6]\,\cite{2014ApJ...781....1L}; [7]\,\cite{2018ApJ...865..107T}; [8]\,\cite{2015A&A...581A..86M}; [9]\,\cite{2022A&A...665A.125R}
    
    }
\end{table*}

\begin{table*}[h]
    \centering
    \small
    \caption{General properties of the prompt emission of the high-redshift GRBs.}
    \label{tab:grb_prompt}
    \begin{tabular}{llllll}
    \hline\hline\\[-2ex]
        Identifier  & T$_{90}$ & E$_{peak}$ & Fluence & E$_{iso}$ & Ref. \\
                    & [s]      & [keV]      & [$\times$10$^{-7}$\,erg/cm$^2$] &  [$\times$10$^{52}$\,erg]     & \\
         \\
    \hline\\[-2ex]
        GRB 050904A & 225$\pm$10   & $>$150       & 54$\pm$2    & 34          & [1,2]   \\
        GRB 080913A & 8$\pm$1      & 93$\pm$56     & 5.6$\pm$0.6 & 7           & [3]     \\
        GRB 090429B & 5.5$\pm$1    & 42$\pm$5.6    & 3.1$\pm$0.3 & 3.5         & [4,5]   \\
        GRB 090709A & 89$\pm$3     & 439$\pm$58    & 257$\pm$3   & 31          & [6,7,8] \\
        GRB 100205A & 26$\pm$8     & ...           & 4.0$\pm$0.7 & ...         & [9]     \\
        GRB 120521C & 26.7$\pm$4.4 & ...           & 11$\pm$1    & 19$\pm$8    & [10]    \\
        GRB 120923A & 27.2$\pm$3   & 44$\pm$11     & 3.2$\pm$0.8 & 4.8$\pm$1.6 & [11,12] \\
        GRB 140515A & 23.4$\pm$2.1 & 51$\pm$15     & 5.9$\pm$0.6 & 5.8$\pm$0.6 & [13,14] \\
        GRB 210905A & 870          & 144$\pm$56  & 182$\pm$29  & 127$\pm$20  & [15]    \\
    \hline\\[-2ex]
    \end{tabular}
    
    \tablebib{
    [1]\,\cite{2005GCN..3938....1S}; [2]\cite{2005A&A...443L...1T}; [3]\,\cite{2009ApJ...693.1610G}; [4]\,\cite{2009GCN..9290....1S}; [5]\,\cite{2011ApJ...736....7C}; [6]\,\cite{2009GCN..9640....1S}; [7]\,\cite{2009GCN..9653....1O}; [8]\,\cite{2010AJ....140..224C}; [9]\,\cite{2010GCN.10371....1S}; [10]\,\cite{2012GCN.13333....1M}; [11]\,\cite{2012GCN.13807....1M}; [12]\,\cite{2018ApJ...865..107T}; [13]\,\cite{2014GCN.16284....1S}; [14]\,\cite{2015A&A...581A..86M}; [15]\,\cite{2022A&A...665A.125R} 
    }
\end{table*}

\begin{figure}
    \centering
    \includegraphics[width=.90\columnwidth]{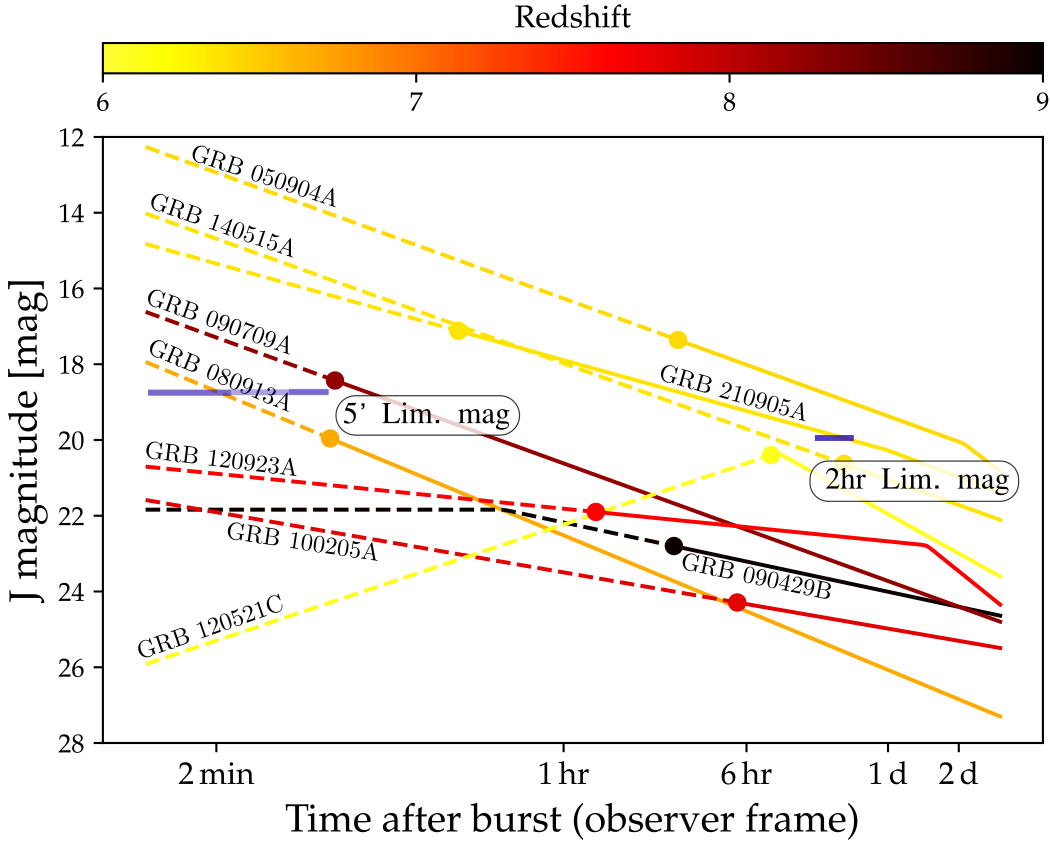}
    \includegraphics[width=.90\columnwidth]{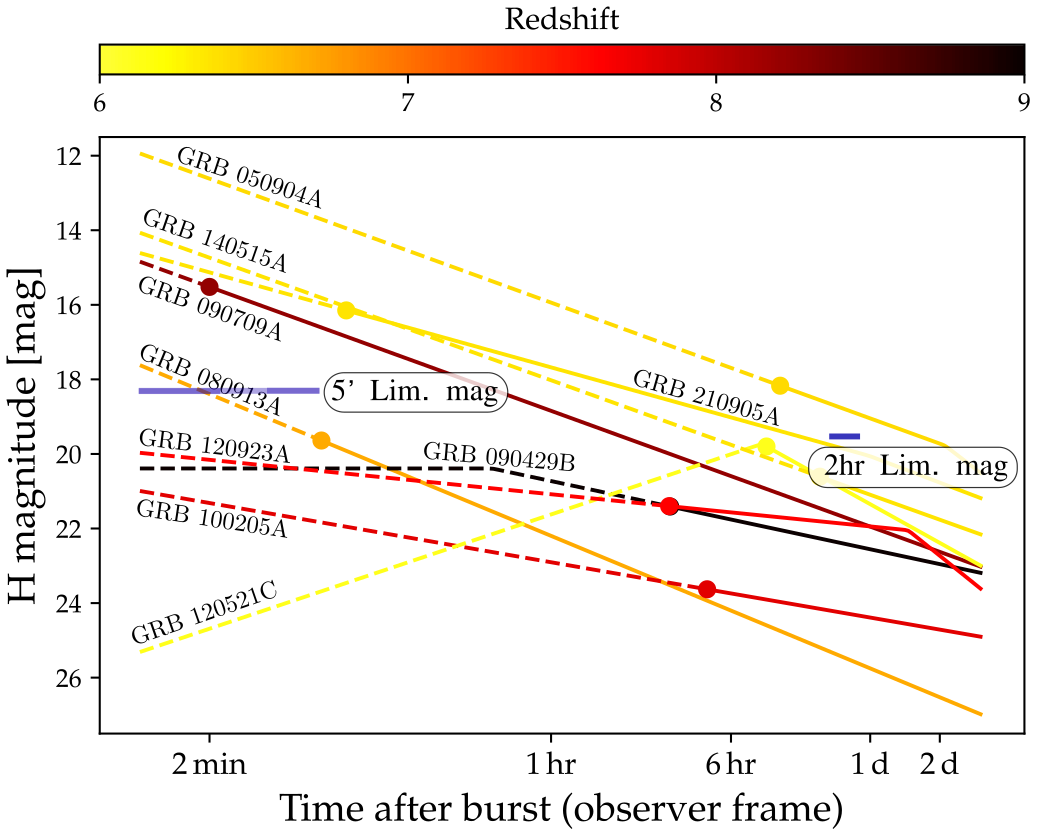}
    \caption{Modelled lightcurves of GRB afterglows in our sample. Filled circles indicate the first photometric measurement after burst found in the literature; dotted lines indicate extrapolated lightcurves in J and H bands. Horizontal blue lines indicate the limiting magnitudes achieved in the early-catch (5' exposures) and late-catch (2\,hr exposures) scenarios (see Sect.\,\ref{sect:results}).}
    \label{fig:GRBs}
\end{figure}

\begin{figure}
    \centering
    \includegraphics[width=.85\columnwidth]{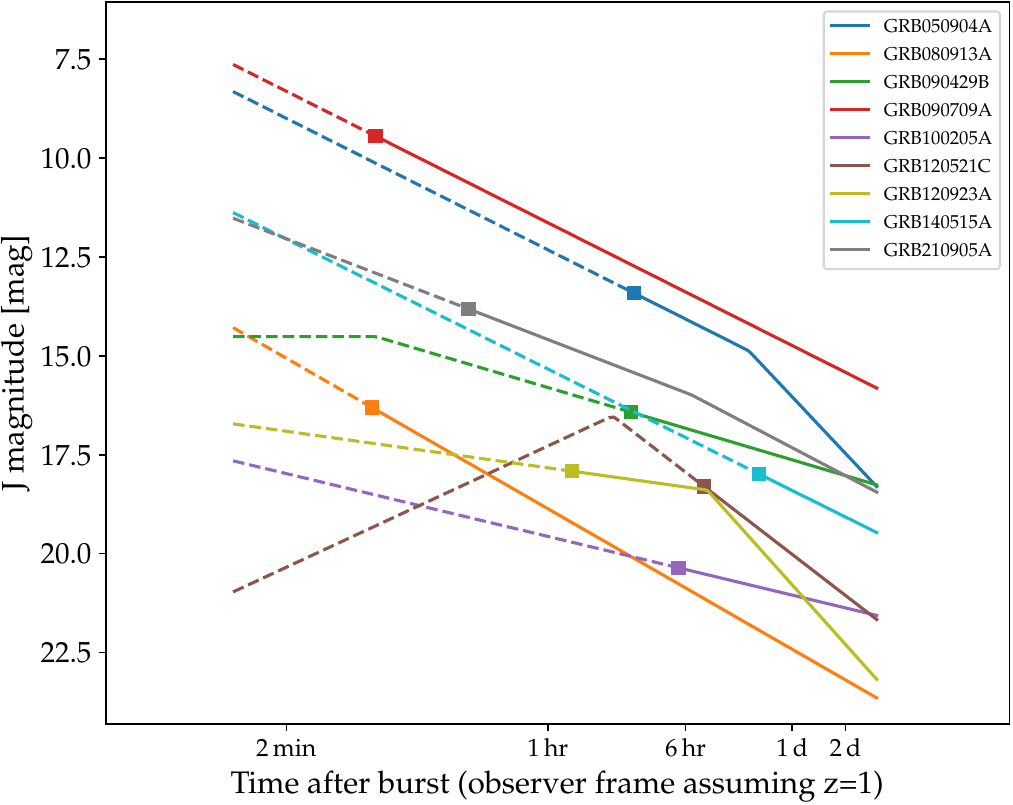}
    \includegraphics[width=.85\columnwidth]{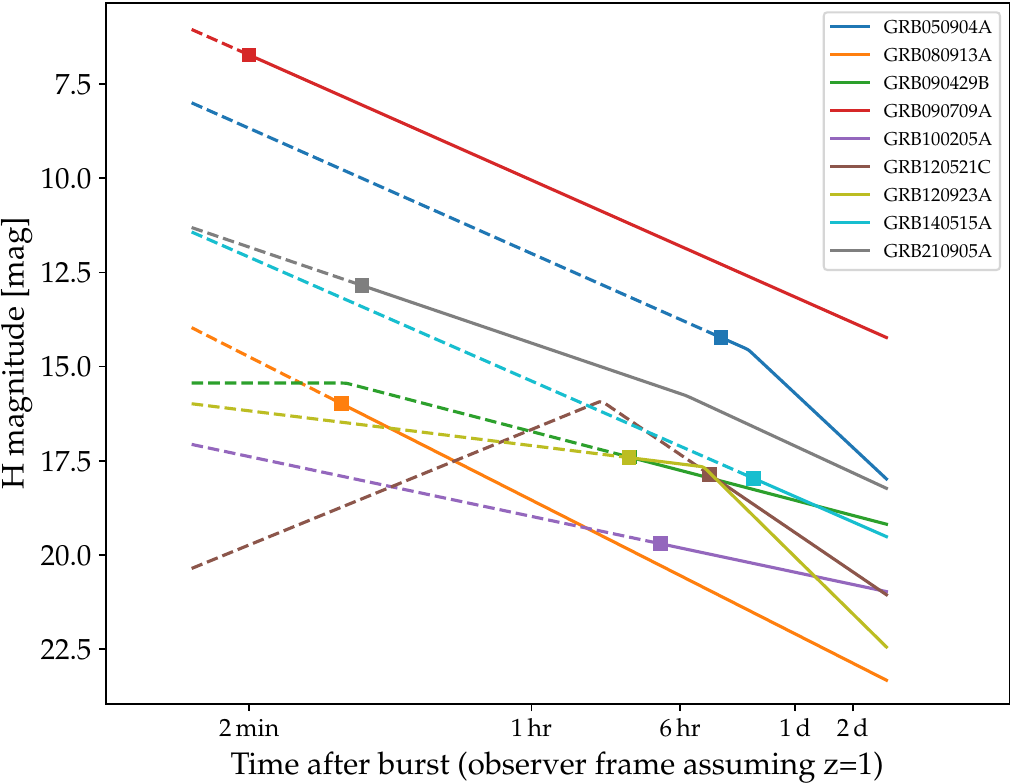}
    \caption{Reconstructed lightcurves of GRB afterglows in our sample, assuming z=1. Squares indicate the first photometric measurement after burst found in the literature; dotted lines thus indicate extrapolated lightcurves in J and H bands.}
    \label{fig:GRBsz1}
\end{figure}


\section{Results}\label{sect:results}

\begin{table*}[h]
    \caption{Detectability of GRBs in the proposed scenarios.}
    \label{tab:grb_detection}
    \centering
    \small
    \begin{tabular}{llllllrr}
    \hline\hline\\[-2ex]
        Identifier  & Redshift  & \multicolumn{2}{c}{Early-catch (1')}  & \multicolumn{2}{c}{Late-catch (12\,hrs)}  & \multicolumn{2}{c}{Maximum redshift} \\
                    &           &  J band   &  H band           & J band   &  H band          & J band  & H band \\
                    &           &  (SNR)    & (SNR)             & (SNR)    & (SNR)            & (3$\sigma$, 10$\sigma$) & (3$\sigma$, 10$\sigma$)\\
    \hline\\[-2ex]
        GRB 050904A & 6.39  & \textbf{174}  & \textbf{321}   & \textbf{8}    & \textbf{7.5} &  9.6,  9.4  & 13.3, 13.1 \\
        GRB 080913A & 6.695 & 0.7  & 1.3                     & <0.1  & <0.1  &  1.2,  0.5  &  1.9,  0.8 \\
        GRB 090429B & 9.4   & 0.2  & 0.5                     & <0.1  & 0.2   &  6.0,  2.8  &  2.5,  1.3 \\
        GRB 090709A & 8.5   & \textbf{3.2}  &\textbf{22}           & <0.1  & 0.5   &  8.4,  3.1  & 12.6  11.5 \\
        GRB 100205A & <8    & <0.1 & <0.1                    & <0.1  & <0.1  &  0.7,  0.3  &  0.8,  0.4 \\
        GRB 120521C & 6.0   & <0.1 & <0.1                    & 0.9   & 1.2   &  0.7,  0.3  &  0.5,  0.4 \\
        GRB 120923A & 7.8   & 0.3  & 0.5                     & 0.3   & 0.4   &  1.5,  0.7  &  1.9,  0.9 \\
        GRB 140515A & 6.32  & \textbf{35}   & \textbf{46}    & 1.7   & 1.1   & 9.3,  8.8   &  12.9, 12.1 \\
        GRB 210905A & 6.318 & \textbf{26}   & \textbf{38}    & \textbf{3.2}  & 2.6   &  9.2,  8.6 &  12.7, 11.8 \\

    \hline\\[-2ex]
    \end{tabular}
    \tablefoot{The values are obtained for observations blocks of 3$\times$100\,s; cases where the afterglow is above the detection threshold (SNR$\geq$3) are indicated in bold.}
\end{table*}

\subsection{CAGIRE detection in the early-catch scenario}
Here we focus on the detectability of GRBs in the best-case scenario, when CAGIRE is able to start observations exactly one minute after the GRB alert. At such an early time after burst, the afterglows are fading very quickly. In our case, if an event is already fainter than the detection limit for 3x100\,s exposures in each filter, increasing the number of exposures is unlikely to allow CAGIRE to catch up with the decreasing flux of the afterglow. Indeed, the decaying rate in the early-catch scenario is such that the the afterglows will typically dim by 3 to 4 magnitudes during the first hour (decaying index $\alpha\,\sim$1); in that time, the increase in limiting magnitude is only about 0.7 mag.

We also note that in such a case, CAGIRE will likely be the first facility to observe in nIR; without prior knowledge of spectrophotometric data coming from other follow-up facilities worldwide, the best course of action is to start the observing blocks with the H band filter. Indeed, while the H band limiting magnitude is slightly worse than in J band, the latter will suffer from the Lyman break at a lower redshift. By starting observations in the J band we would lose the advantage of COLIBRI's fast reacting time for bursts at z$>$9. Starting early observations in H band rather than J minimises the chances of missing a high-redshift or obscured event, which is in line with CAGIRE's main science goal of detecting the afterglows of far away and obscured GRBs.

Out of the 9 GRBs we simulated, CAGIRE is able to detect GRB 050904A, GRB 090709A, GRB 140515A and GRB 210905A  with three exposures of 100\,s in each band. The non-detected GRBs are located at redshifts $>$7.8, with the exception of GRB 120521C which is located at z=6.03. This event is peculiar though, as the model of its early light curve indicates a rapidly rising source (hence the negative exponent in Table\,\ref{tab:grb_prop}) before the break at around 8\,hr. This results in a very faint source when extrapolating to earlier observation times. Although, even if caught at its peak infrared luminosity (J=19.7, H=20.35), it would still be too faint to be reachable by CAGIRE.

Finally, we have also computed the SNR reached in the case of CAGIRE starting observations 2\,min after burst instead of 1\,min to asses the impact of 
slight delays in the pointing time at such an early phase of the afterglow. Across our sample, we lose 20\% of SNR in the H band and about 8\% in the J band; this difference is explained by the H observations taking place earlier, when the decay rate of afterglows are very steep. In this particular case, we cannot detect GRB 090709A in the J band any more, as its SNR passes right below 3.

\subsection{CAGIRE detection in the late-catch scenario}
In the case a GRB is detected by \textit{SVOM}/ECLAIRs when the sun is rising in San Pedro Mart\'ir, COLIBRI will not be able to point to the target before several hours when the sky is dark again: in the worst-case scenario, this would be an average of 12 hours after trigger; of course, any intermediate cases are possible depending on the time of day when the burst happens. In this worst-case scenario, the afterglow magnitude is much fainter than at early times, but the fading rate is also much slower. For typical decay indices of $\alpha\,\sim$1, the afterglows dim by around 0.1 magnitudes between 12 and 13 hours after burst; in this case, an afterglow that is less than 0.6 magnitudes below the detection limit for 300\,s exposures can be detected by increasing the number of exposures. Hence, the best observing strategy here is to take the deepest images possible, as the increase of SNR with exposure time is potentially able to catch up with the fading afterglows. This is the case for GRB 210905A, for which 2\,hr exposures in the late-catch scenario allow us to detect it just above SNR=3 in the J band, when only 300\,s would not have been enough to pass that threshold.

In the case of the 9 listed GRBs in this study, 2\,hr exposures (72$\times$100\,s) are able to reach the magnitude of the two brightest afterglows, GRB 040905A and GRB 210905A. For any other undetected afterglow, such exposures will provide upper limits of J$\sim$20 and H$\sim$19.5 at three sigma.

However, as discussed in Section\,\ref{sect:cagire}, we can increase the magnitude limit in J band by performing 300\,s individual exposures instead of 100\,s. This is possible in this late-catch scenario as this doesn't impede on the H band exposures and the increase in SNR with exposure time is usually faster than the decaying rate of the afterglows. For 24$\times$300\,s exposures, we reach a magnitude limit of J=20.47 at SNR=3, which is 0.5 magnitudes deeper than 72$\times$100\,s exposures. This translates into a 57\% increase in SNR for J band images. In our list of GRBs, the afterglow of GRB 140515A thus reaches a SNR of 2.7 (instead of 1.7), just below the detection threshold. Hence in the late-catch scenario, there are only advantages in increasing the individual J band exposures from 100\,s to 300\,s. This will maximise the chances of detecting the GRB counterparts, and at worse will give more constraints on the flux upper limit.

\subsection{Detecting very high redshift events with CAGIRE}\label{subsect:highz}

Beyond redshift 6, the impact of luminosity distance on the flux is marginal, and is almost offset by the effect of time dilation since the afterglows are effectively caught at an earlier time after the burst in their rest frame; the main factor governing the loss of flux is the sharp decrease below the Lyman\,$\alpha$ wavelength. Hence, for intrinsically bright events, CAGIRE is not limited by its specifications nor by the collecting power of COLIBRI, but by the extreme obscuration caused by the Lyman break reaching the J and H photometric bands at high redshifts.

This results in CAGIRE being able to detect bright events such as GRB 050904A, GRB 090709A, GRB 140515A and GRB 210905A up to redshifts of 9.6, 8.4, 9.3 and 9.2 respectively in J band, and redshifts of 13.3, 12.6, 12.9 and 12.7 respectively in H band at SNR=3 in the early-catch scenario (Table\,\ref{tab:grb_detection}). In a late-catch scenario, we are not able to detect the GRBs at significantly greater redshifts, since their detection at their actual redshifts are already marginal. This reinforces the fact that the uniqueness of CAGIRE resides in the combination of a near-infrared imager with a fast-slewing robotic telescope that is able to work in synergy with SVOM/ECLAIRs as early as two minutes after burst.

\section{Discussion}\label{sect:discussion}

\subsection{Locating GRB afterglows at early times}
Locating transient sources within astronomical images necessitates dedicated pipelines, that are yet to be developed for CAGIRE. The SNR of the transient will have an impact on the ability of those pipelines to securely identify it; a SNR greater or equal than 10 should allow for a reliable detection independently of the algorithm used. All of the detected GRBs we simulated fulfil this criteria for the early-catch scenario (except GRB 090709A in J band), and none of them do for late-catch scenario. This means catching GRB afterglows as early as 1 minute after burst is crucial to provide the location of the counterpart to the community and ensure efficient follow-up by larger facilities, especially for spectroscopy where good accuracy is needed to position the slits. 

Sampling the early lightcurves of GRB afterglows also has an intrinsic benefit of probing an epoch rarely observed in the near-infrared domain. COLIBRI and CAGIRE have the potential to produce a database of early lightcurves where the evolution of flux may significantly differ from the decaying powerlaw model, caveat which is assumed in our study to extrapolate reconstructed lightcurves at times earlier than 1\,hr.

\subsection{Lyman alpha flux deficit evolution at high redshifts}
The only GRB in our sample that has a magnitude measurement landing within the Lyman forest is GRB 090429B. At redshift 9.4, the J band is heavily impacted by the Lyman forest. The model we use to simulate this deficit, given by \cite{1993ApJ...418..601Z}, is based on quasi-stellar objects up to redshift 4. When artificially putting GRB 090429B at a closer distance, this model seems to slightly overestimate the flux loss, such as the J band magnitude becomes brighter than the H band. Hence, it is likely that by observing the afterglows of far away GRBs, it will be possible to better constrain the evolution of the effects of Lyman\,$\alpha$ absorption on a cosmological scale. Along with allowing for a better reconstruction of the intrinsic lightcurves and spectra of far away GRBs, this would be a way to probe the intergalactic medium and the formation of early galaxies.

\subsection{CAGIRE and the general population of GRBs}
\begin{figure}
    \centering
    \includegraphics[width=0.9\columnwidth]{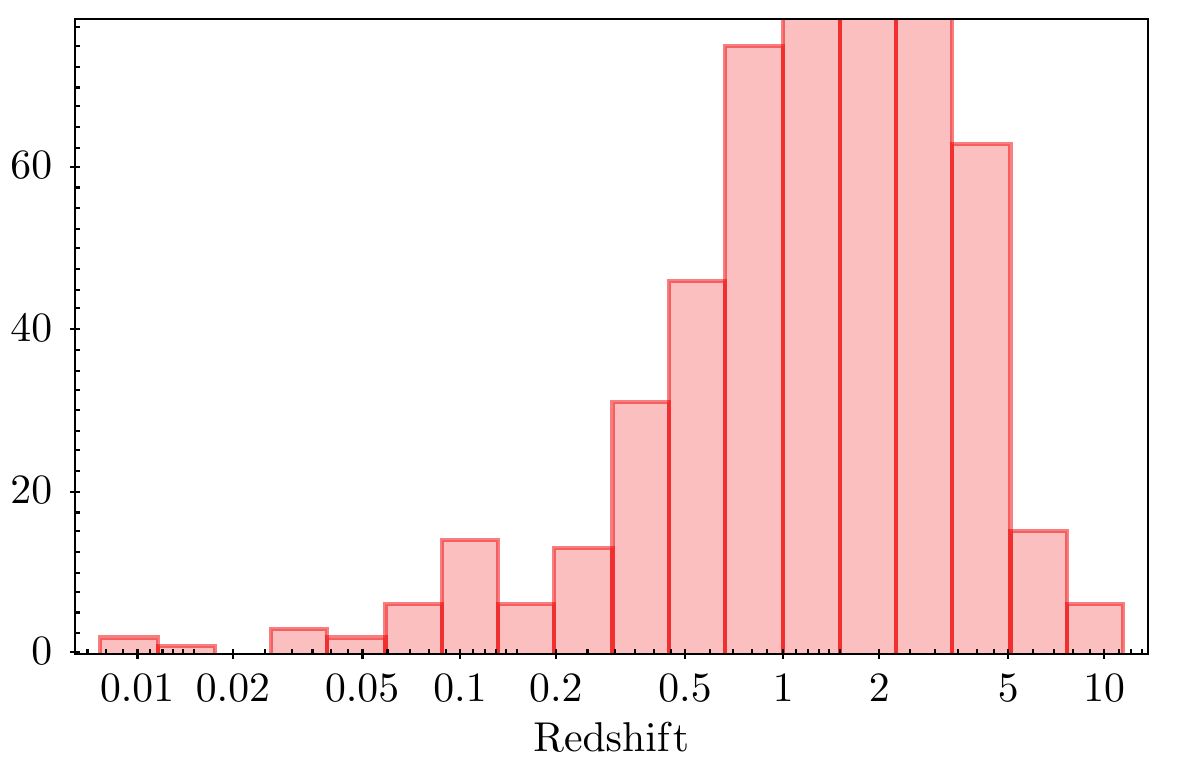}
    \caption{Distribution of measured redshifts in GRBs according to GRBweb.}
    \label{fig:grbz}
\end{figure}

The main point of this paper is about estimating the performances of CAGIRE in the scope of its scientific goals, i.e. the detection of obscured/far away GRB afterglows. However, according to GRBWeb\footnote{\url{https://user-web.icecube.wisc.edu/~grbweb_public/index.html}}, 98\% of GRBs with an identified redshift lie at z$<$6 (see Figure\,\ref{fig:grbz} constructed from the data available in GRBWeb). Although the true distribution of GRB redshift is unknown, it is likely that high redshift GRBs will constitute rare events for CAGIRE. Hence, the camera will follow-up many afterglows located closer, and we estimate here how CAGIRE would fare in this scenario.

We combine two collections of near infrared lightcurves of past GRB afterglows. The first is a compilation by Steve Schulze of composite Rc-band lightcurves between 1997 and 2013, based on the works of \cite{2006ApJ...641..993K,2010ApJ...720.1513K,2011ApJ...734...96K,2012A&A...548A.101N}. It consists of 176 lightcurves with varying coverage after burst and time sampling. The second compilation is an ongoing project by Damien Turpin (priv. communication) of retrieving photometric measurements of GRB afterglows directly from GCN circulars and the literature. It compiles 2567 near-infrared photometric measurements in J and H bands on a total of 279 GRB afterglows. So far, the total sample we work with consists of 364 unique afterglows. We also acknowledge the work of \cite{2024MNRAS.533.4023D} who compiled similar data on 535 GRBs, as well as the GRBphot database mentioned in \cite{2020SPIE11452E..18B} and \cite{2024A&A...686A..56K}; however both photometric databases were not available at the time of the data compilation. Since then,  GRBphot has been released on Vizier and we used it to supplement our database. We found 246 photometric measurements in either R$_c$, J or H bands on 24 individual bursts. Our database already contained information on 17 of these bursts, meaning we can work with an extra 7 events. Two of them have an insufficient number of datapoints in their lightcurves for us to use them in our simulations, hence we are able to supplement our database with 5 extra GRB lightcurves, for a total of 369.

Rc magnitudes are converted to J and H magnitudes using a powerlaw with $\beta$=1, which is the average value found in our sample of far away GRBs (though it may vary on across the population of all GRBs). To retrieve the magnitudes of the afterglows at t=1\,min (early-catch) and t=12\,hrs (late-catch), we first check if there are data points within 10s--20\,min (early-catch) and 2\,hrs--5\,days (late-catch); if so, we assume a decay index $\alpha$=1; otherwise, we consider the data points to be too far in time to reasonably extrapolate the lightcurves and do not use them.


In the early-catch scenario, we count 142 (150) J band (H band) measurements that we can use. Of them, 12 (13) have an SNR below 3, meaning they are not detected in the J (H) band. Out of the 130 (137) afterglows detected at SNR greater than 3, 111 (118) have a SNR greater than 10 in the J (H) band. This results in an overall detection success rate greater than 90\% in the early-catch scenario in both J and H bands.


In the late-catch scenario, we count 327 (312) J band (H band) measurements that we can use. Of them, 144 (191) have an SNR below 3, meaning they are not detected in the J (H) band. Out of the 183 (121) afterglows detected at SNR greater than 3, 89 (54) have a SNR greater than 10 in the J (H) band. This results in an overall detection success rate of 56\% in J and 39\% in H bands in the late-catch scenario. Using longer individual exposures (300\,s) in the J band provides a slightly better detection success rate of 60\% (+13 detections). Again, this is not applicable in the H band as the sky contribution becomes too important past 100\,s exposures. We conclude that for a more generic population of GRB afterglows, CAGIRE will be particularly efficient at sampling the early lightcurves in both J and H bands, and will still be a great source of photometric measurements many hours after the bursts occur.

\subsection{Observatory efficiency}
It is difficult to predict the number and nature of GRB afterglows that will effectively be detected by CAGIRE. The population of GRBs seen by SVOM is indeed hard to anticipate because the low energy threshold of ECLAIRs (helping towards detecting high-redshift GRBs, \citel{2024A&A...685A.163L}), the fast-reaction time of COLIBRI and the nIR capabilities of CAGIRE open a new parameter space in transient sky astronomy.

We can however interpret the detectability numbers we provide in the previous section in light of known data about the San Pedro Mart\'ir observatory site and requirements. According to \cite{2012MNRAS.420.1273C}, we can expect an average clear sky fraction greater than 80\%\, at the observatory location. The Functional Performance Requirements document for COLIBRI also specifies that the observatory should operate 90\%\, of the nights, the predicted time loss being due to both software and hardware issues. For reference, the contribution of CAGIRE's hardware problems amounts to 2\%, or about 7 nights per year. These observational restrictions should not significantly impact the ability of CAGIRE to detect far away and obscured GRB afterglows.

\section{Conclusion}\label{sect:conclusion}
Once CAGIRE is operational on the COLIBRI telescope (Q2-Q3 2025), the ground segment of \textit{SVOM} will have a unique instrument for the follow-up of GRB afterglows. The nIR camera will have the capability of rapid slewing after burst triggers issued from \textit{SVOM}/ECLAIRs (T-T$_0\sim$1\,min) and identify a counterpart thanks to CAGIRE's field of view covering ECLAIRs error boxes in a single exposure. We have simulated nIR images output of CAGIRE using an end-to-end simulator in order to realistically assess the performances of the instrument on known GRBs. We modelled the lightcurves of their afterglows, and produced artificially redshifted events to explore the best capabilities of CAGIRE; out of the 9 GRBs with z$\geq$6 with sufficient photometry coverage in the literature to produce lightcurve models, CAGIRE is able to detect the afterglow of the 4 brightest events. Similar events happening at even greater redshifts can be detected up to z=9.6 (J band) and z=13.3 (H band). It is the combination of being able to produce images very early after the high-energy triggers with the infrared pushing back the Lyman\,$\alpha$ break limit that gives CAGIRE the potential to identify the counterparts of the farthest GRBs to date (z$>$9.4). On the general population of GRBs, CAGIRE will perform very well, being able to catch more than 90\% of afterglow counterparts at early times.

Sampling the early lightcurves of far away GRBs provides a unique opportunity to probe the very first instants of their afterglows thanks to time dilation effects. Less than 20\% of the GRBs in our sample have photometry available at times earlier than 2 minutes in near infrared, most of which are reconstructed from multi-band data. These early epochs are likely to differ from a typical decaying powerlaw and CAGIRE will be capable of systematically providing early photometry to constrain their evolution.

Beyond redshift 6, CAGIRE will also be able to see GRBs issued from population III massive stars that started the reionization epoch. This may help resolving the bias between star formation rates inferred by the currently observed occurrence of GRBs \citep{2013A&A...556A..90W} and the models using the interstellar medium properties of high-redshift galaxies (see e.g. \citel{2022A&A...663A.172M}). The evolution in the composition of the intergalactic medium can also be probed through the shape of the Lyman forest, thus allowing to put observational constraints on cosmological parameters \citep{2021JCAP...10..077S}. Thus, the potential scientific results brought by CAGIRE in the scope of the SVOM mission reach many domains of astrophysics, even further expanded as COLIBRI, operating as an observatory independently of SVOM, will be available for the wider community of astronomers.

\begin{acknowledgements}
F. Fortin is grateful for the funding from CNES. We thank Lynred, ESA and the LabEx FOCUS for making the ALFA detector available for CAGIRE. We also thank Steve Schulze for making their database public on GitHub (\url{https://github.com/steveschulze/kann_optical_afterglows}).

\end{acknowledgements}

\bibliographystyle{aa}
\bibliography{references}

\end{document}